# Functionalization of structural and electronic properties of multiferroic $SrTiO_3$ thin films

**Zdeněk Jansa**[1)]; Lucie Prušáková[1)]; Štěpánka Jansová[1)]; Pavel Calta[1)]; Pavol Šutta[2)]; Ján Minár[1)]

[1)] *West Bohemia University, Pilsne, Czech Republic*

[2)] *Liptovský Mikuláš, Slovak Republic*

Abstract

This study examines the application of transition metal-doped $SrTiO_3$ in photovoltaic technologies, such as photocatalysis. The core objective is to evaluate how different dopants influence the structural and electronic characteristics of the well-known perovskite, $SrTiO_3$ (STO). By incorporating dopants, particularly transition metals, the material's physical properties can be enhanced by addressing limitations such as the large gap in the valence band. This study aims to determine the impact of these metals on factors like crystallite size, internal stress levels, electron-hole pair distribution in the valence band, and the shift in the electromagnetic spectrum toward the visible range. The primary focus is on assessing nickel's (Ni) influence on these properties, with additional investigation into the effects of yttrium (Y) and iron (Fe).

Several experimental methods were employed to analyze the structural and electronic properties of $SrTiO_3$. The same procedures were applied consistently across all samples, with the sole exception being the high-temperature XRD experiments, as described in the text. The techniques used include magnetron pulse deposition for sample preparation, followed by X-ray diffraction (XRD), scanning electron microscopy (SEM), and X-ray photoelectron spectroscopy (XPS). After conducting the experiments, the collected data were evaluated and used to guide subsequent steps. Finally, all data were consolidated and analyzed comprehensively.

## 1. Introduction

The minerals known as Perovskites were first identified in the Ural Mountains in 1839, with the original perovskite having the chemical formula $CaTiO_3$. This discovery led to the naming of a group of compounds, now called Perovskites, which includes hundreds of similar compounds today. These newer compounds typically follow a chemical structure based on or closely resembling the original formula, $ABX_3$. The most abundant solid mineral on Earth, Bridgmanite, belongs to this group, with a chemical formula of $(Fe,Mg)SiO_3$.

Perovskite compounds are ionic in nature. In the $ABX_3$ formula, A and B represent two cations, with A being larger and B smaller, while X represents a small anion. To maintain overall neutrality, the charges of the ions (qa, qb, and qx) must satisfy the equation:

$$q_a+q_b=-3q_x$$

Transition metal oxides with a perovskite structure have gained significant attention in recent decades due to their desirable properties, which make them suitable for various industries. These properties include strong negative magnetoresistance, multiferroicity (exhibiting ferroelectric, thermoelectric, pyroelectric, dielectric behavior), and excellent optical qualities. As a result, perovskites are utilized in capacitors, memory devices, tunable microwave components, displays, piezoelectric devices, sensors, actuators, transducers, and wireless communication systems.[1]

In the energy sector, perovskite materials have shown promising applications. Recent experiments demonstrate that, with certain modifications, perovskite oxides can harness direct sunlight and be used in photocatalytic processes. The original strontium titanate oxide ($SrTiO_3$ or STO) can only absorb UV radiation, remaining transparent to visible light due to its wide band gap (3.2–3.25 eV at room temperature). Studies have indicated that doping these oxides with transition metals (TM) can alter the valence and conduction bands, reducing the band gap by introducing new energy levels.

Nickel, one of the transition metals, is suggested by research to exist as Ni2+ in the cubic STO structure, replacing Ti4+ ions. The absorption edge of STO relative to NiO shifts by 1.1 eV when doped with nickel. This study focuses on examining the structural differences in STO samples prepared via magnetron deposition with varying levels of nickel dopant. Additionally, the impact of the preparation environment (whether vacuum or with oxygen flow during deposition) on the structure was investigated. The experiment aimed to verify the formation of oxygen vacancies within the STO structure, which could potentially lead to photocatalytic aktivity. [2,3,4]

## 2. Samples and Methods

All samples were produced in a magnetron deposition chamber using the TF 600 BOC Edward deposition system, which was employed for the growth of all STO thin films in the study. This setup allows for thin film preparation through magnetron sputtering, utilizing two magnetrons—one powered by a high-frequency (HF) source (13.56 MHz) and the other by a DC source. Two batches of samples were prepared in the chamber. For the first batch, the chamber was evacuated to a base pressure of $2x10^{-4}$ Pa and then filled with argon as the working gas. Strontium titanate (STO) layers without dopants were initially deposited to serve as a reference for analyzing the impact of dopant elements on the layer structure. A 99.9% pure sintered STO target on a magnetron powered by a high-frequency source was used for these depositions. Polished silicon substrates with (111) orientation were cleaned using acetone and isopropyl alcohol in an ultrasonic bath for 10 minutes each before deposition. Next, STO layers doped with nickel pellets were deposited using a similar process. Subsequently, yttria-doped STO thin films were created using two targets, one for STO on a high-frequency magnetron, and the other for yttria on a DC-powered magnetron. Finally, iron-doped STO layers were deposited using a process identical to the nickel-doping method, with iron pellets placed on an STO target powered by a high-frequency source.[5,6,7,8]

The experimental procedures were identical for both series of samples. After fabrication, all samples underwent X-ray phase analysis. X-ray diffraction (XRD) measurements were performed using a Panalytical X'Pert automatic powder diffractometer. Measurements were taken in two geometries available on the machine's goniometer. The first was symmetric $\theta$-$\theta$ geometry, while the second was an asymmetric $\omega$-$2\theta$ geometry with a fixed X-ray tube angle of $\omega = 0.5°$. This method, known as Grazing Incidence Diffraction (GID), ensures shallow penetration of the primary beam into the sample while irradiating a large surface area. Both geometries were used in the 2θ range of 20° to 75°. Initially, the samples were remeasured by GID in their as-deposited state, then subjected to an in-situ experiment in a high-temperature chamber, reaching up to 900°C. Given the time constraints of the experiment, the samples were measured in the high-temperature chamber using the θ-θ geometry. After removal from the chamber, the samples were remeasured using GID, and the data were subsequently analyzed.

The second experimental technique employed was scanning electron microscopy (SEM). A JSM-7600F electron microscope from JEOL was used to observe the chemical composition, surface morphology, and fractures of the samples.

The third method used was X-ray photoelectron spectroscopy (XPS). These experiments were conducted on selected samples in an ultra-high vacuum working chamber (base pressure ≤ $5\times10^{-8}$ Pa), using a Phoibos 150 hemispherical analyzer. A magnesium and aluminum X-ray tube was employed as the photon source throughout the experiments. The experiments followed this procedure: Broad spectra were measured at an Epass pass energy of 50 eV, with an energy step of 0.5 eV and a dwell time of 96 ms. Detailed O 1s and C 1s spectra were obtained at 30 eV with a 0.1 eV step and a dwell time of 198 ms. For better resolution of the Sr 3d, Ti 2p, Fe 2p, Ni 2p, and Y 3d peaks, a 0.05 eV step was used. The number of scans for dopant peaks was increased from 20 to 50 to improve accuracy due to the small amount of dopants in the thin film structure.

## 3. Results and discussions

### XRD

The first measurement of all samples in the as-deposited state was performed using asymmetric geometry. The measurement range was set from 20 to 70° 2θ. The measurement conditions were identical for all samples, i.e. the measurement was performed at room temperature, with a time set to one measurement step of 30 sec (step 0.06° 2θ). The samples were of uniform size 10x10 mm and all were deposited on Si substrate.

Figure 1 shows an overlay of the diffractograms of all samples in the as-deposited state. As can be seen, all the samples contain an amorphous phase, which appears as broad lines in the recordings. The STO standard with the designation 00-040-1500 from the International Centre for Diffraction Data (ICDD) database, which has a cubic grid, has been inserted into this plot. This is the same crystallographic lattice as the pure STO, however, due to nickel doping, there are changes in the inter-plane distances caused by deformations of the cubic lattice and thus shifts in the diffraction lines.

The recordings show an interesting effect of the influence of the added dopant. This effect is seen in the partial incipient crystallization of the originally amorphous phase. As the Ni content increases, the diffraction patterns of the STO crystallographic planes are more accurately identified. At the same time, the $TiO_2$ phase, which is an accompanying minority phase formed during STO deposition, appears in the diffractograms. This phase was already present in the evaluation of diffraction records of undoped STO. As can be seen in the recordings, sharp diffraction lines belonging to Laue diffraction are present in all samples.

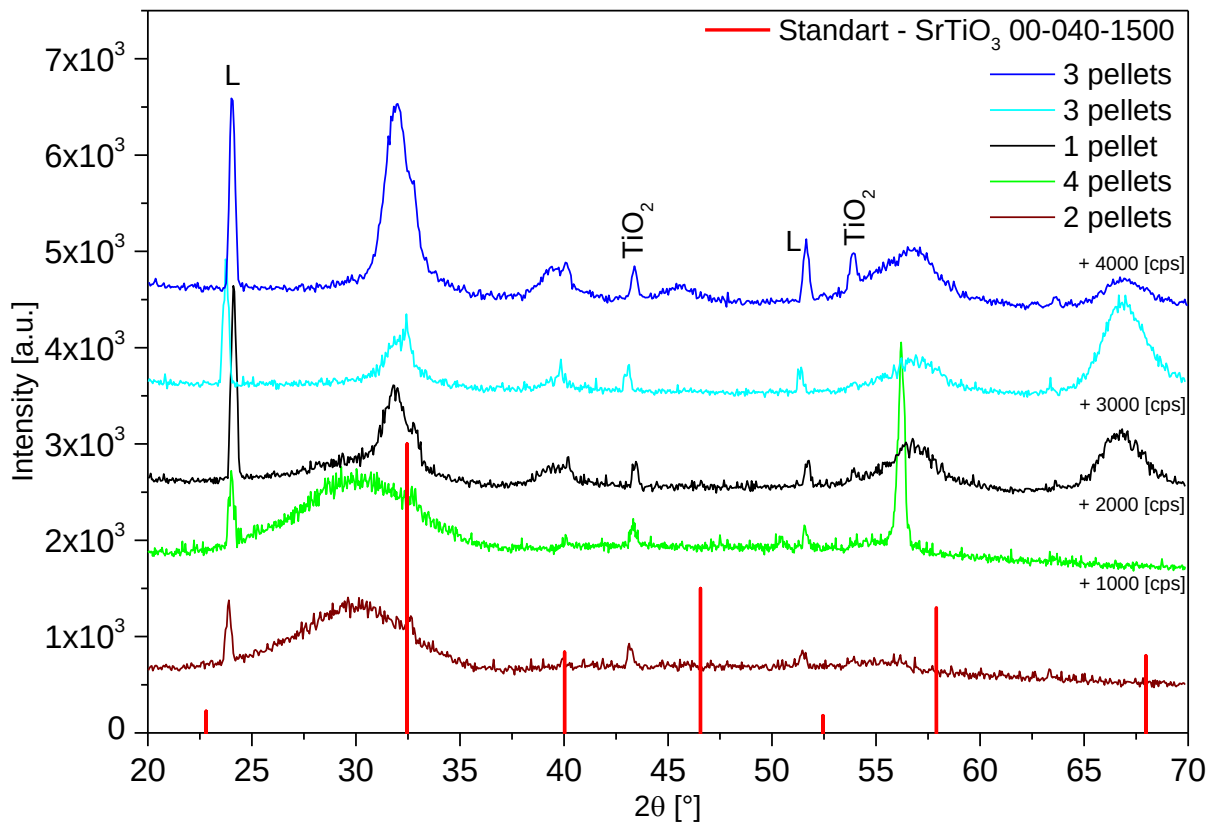

Fig. 1 – Overlay samples STO:$Ni_x$ – as-deposited

After the evaluation of the first part, the experiment continued with annealing in a high temperature chamber. The samples were successively loaded into the HTK 1100 high temperature chamber where the in-situ experiment was performed. During this process, the samples were annealed to 900°C in a high vacuum environment to effect the phase transformation from amorphous to crystalline phase.

The measurement result of the sample with two pellets is presented in Figure 2(a). By tentative evaluation of EDS, it was determined to contain 1.06 at % dopant. The record shows that full crystallization of the initially predominantly amorphous sample occurred. All diffraction lines of the cubic form of STO are present. Analysis of further diffraction lines revealed that the structure also contains minor $Ni(TiO_3)$ phases with a rhombohedral lattice and, as a residue after deposition, $TiO_2$ (anatase) with a tetragonal lattice. The abundance of the $TiO_2$ phase is at the very limit of reliable determination. The $TiO_2$ phase was already present in the thin film structure in the as-deposited state and its amount was reduced by subsequent crystallization of the thin film and rearrangement of the atoms of the individual elements. The $Ni(TiO_3)$ phase was only formed during crystallization in the annealing process and its formation can be explained as a complete substitution of Sr in the original STO lattice associated with a change in the crystal lattice.

Figure 2(b) presents the result of the measurement of a sample doped with 4 Ni pellets, for which the Ni content in the thin film structure was tentatively determined to be 4.43 at %. By diffraction line analysis, it was found that in addition to the major phase of STO, not only the $NiTiO_3$ phase but also the $Sr_2$ and $Sr_4O_4$ phases, i.e. phases that were formed from the substituted strontium released by nickel, appear in the structure. Furthermore, the nickel oxide phase Ni4O4 was present in the structure. The presence of this phase is inappropriate in the structure as it forms metal clusters and reduces the efficiency in the photocatalytic process [9].

The result of phase analysis of the single pellet sample is shown in Figure 2(c). The amount of nickel dopant determined by EDS was 0.81 at %. As can be seen from the diffractogram, such a small amount of dopant (even though it is only 0.25 at % less than the sample with two pellets) caused the intensity of the (220) plane to increase enormously. Even though it is the same direction [110] of the cubic lattice, the increase in the second order plane is a negative effect. This condition may be due to imperfect crystallization, where this quantity hindered the phase transformation process in the high temperature chamber. In fact, the record shows an amorphous hump at the 56° [2θ] position as well as the absence of the STO diffraction plane (100) and a significantly higher diffraction line width at half maximum intensity (FWHM). For example, the difference between the two-pellet and single-pellet samples is 0.4031° [2θ] versus 0.4427° [2θ] for the STO phase and crystallographic plane (110) and 0.502° [2θ] versus 0.6612° [2θ] for the plane (220). Also for this sample, the $Sr_4O_4$, and $NiTiO_3$ phases were present.

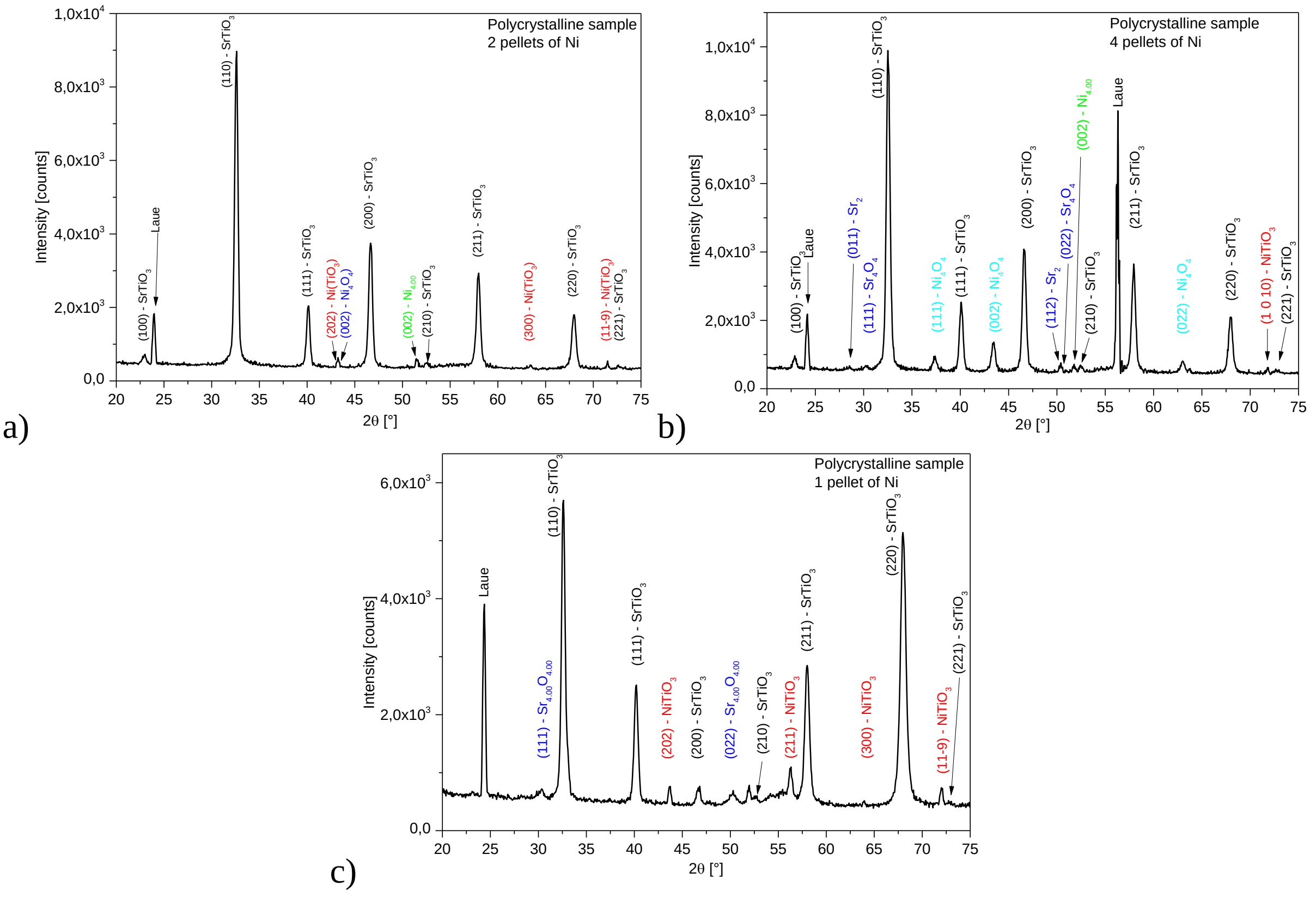

Fig 2 – Diffractograms of samples a) with 2 pellets of Ni, b) with 4 pellets and c) with one pellet. The records shows state ofter heating in HTK and phase transformation from amorphous phase (as-deposited) to crystal phase

A comparison of the diffractograms of all samples is then shown in Figure 3(a). Here it can be seen that the sample doped with one Ni pellet contains the least amount of minority phases. Furthermore, this sample is the closest to the compared standard STO 01-084-0443, including the ratio of the intensities of the individual diffraction lines. For the sample with 4 Ni pellets used, a nickel oxide phase appears in the structure. Samples with one pellet and three pellets are so affected by the amount of dopant that they show an enhancement of the second order diffraction direction [110]. For the sample with three pellets, there is even an inversion of the intensities and the strongest plane is (220).

The effect of the amount of dopant in the structure of the doped STO thin film on the crystallite size is demonstrated in Figure 3(b). In this figure, the strongest diffraction line of the (110) plane is compared. All samples of the first series are shown, including a comparison with undoped STO and the position of this line from STO standard No. 01-084-0443. As illustrated graphically, our doped STO:Nix thin films are all affected by different amounts of dopant in the structure. The diffraction maximum is shifted at different values of the 2θ angle, not only from the standard used, but also (more importantly) from the undoped STO. Both undoped STO samples are shifted 0.09° to the right on the x-axis from the standard used, which means that the interplane distances will be smaller. At the same time, however, the doped STO:Nix must be considered with the undoped STO, which is thus the sample. Thus, this comparison shows that only in the case of the sample with 4 pellets used is compressive stress present, since the maximum of line (110) is shifted to the left. All the other samples are shifted to the right and hence the dopant effect induces tensile stress in these samples. The use of two pellets is best presented in this context.

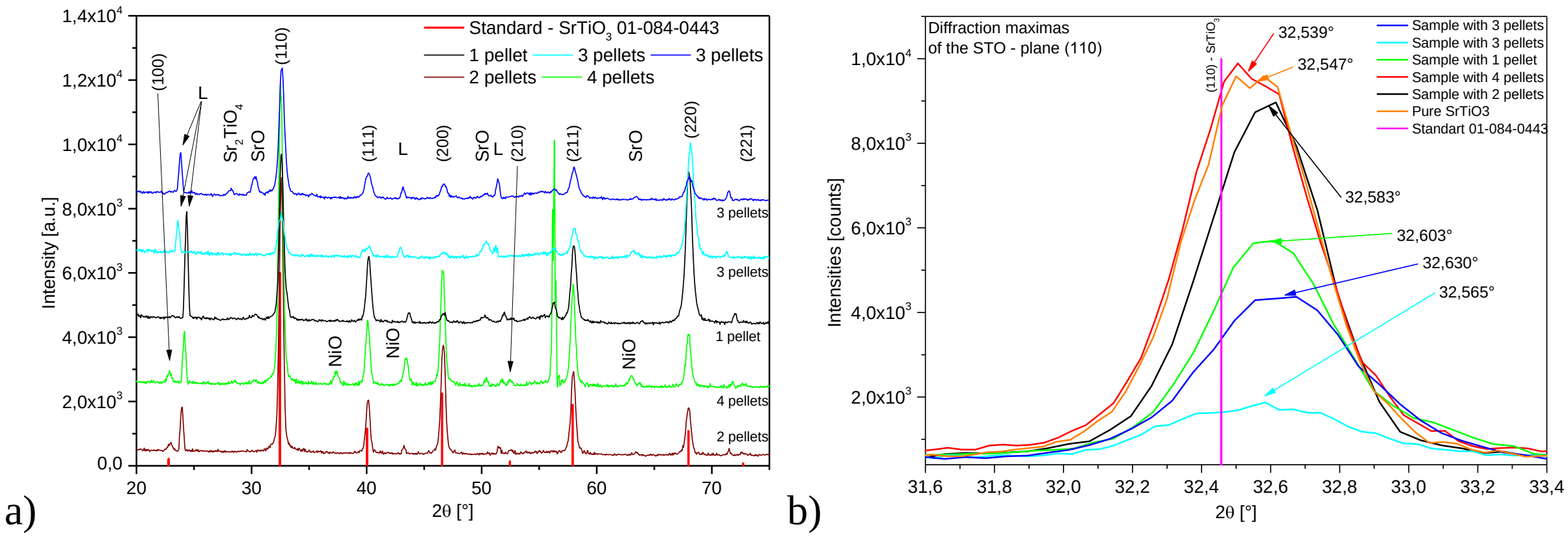

Fig 3 – A comparision of all samples STO:Ni$_x$. Fig 3 a) present overlay of all samples. Fig 3 b) shows a dependence of the amount of nickel dopant on the shift of the maximum of the diffraction line of the STO plane (110).

The second dopant used to improve the structural properties of undoped STO was yttrium.

Several samples were created with different amounts of dopant. This amount was controlled by the power on the DC magnetron and then the amount of dopant was tentatively determined using the EDS method. The amount ranged from 0.12 to 4.29 at. %. When the STO:Y$_x$ samples were measured in the as-deposited state, they were found to be predominantly composed of the amorphous phase, as were the STO:Ni$_x$ samples. Only a few faint diffraction lines, whose shape indicates incipient crystallization, appeared on the diffractograms, belonging to the TiO and $Y_2Ti_2O_7$ phases.

The maximum annealing temperature was adjusted during the high-temperature experiment. The temperature reduction was a two-step process. First, the temperature was reduced to 700°C and then to 500°C. In both cases the same phase transformation from amorphous to crystalline phase was achieved. At the same time, the sample was much less thermally stressed and the thermal stress was lower. At the same time, a difference in some of the accompanying phases was observed, as shown in Figures 4(a) and (b) for the vortex with a dopant amount of 1.39 at% yttrium. In the structure, the STO phase with cubic lattice was identified as the major phase, followed by the $Sr_2TiO_4$ phase with tetragonal lattice and the $YTiO_3$ phase with ortho-rhombic lattice. These are the same phases that were also identified in the identical sample annealed at 700°C, but where an additional phase was identified in addition, see Figure 4(a). Unlike the identical sample annealed to 700°C, no other phase crystallized at the maximum temperature of 500 °C. The diffractograms of the sample include the diffraction lines of the STO 01-089-4934 standard used.

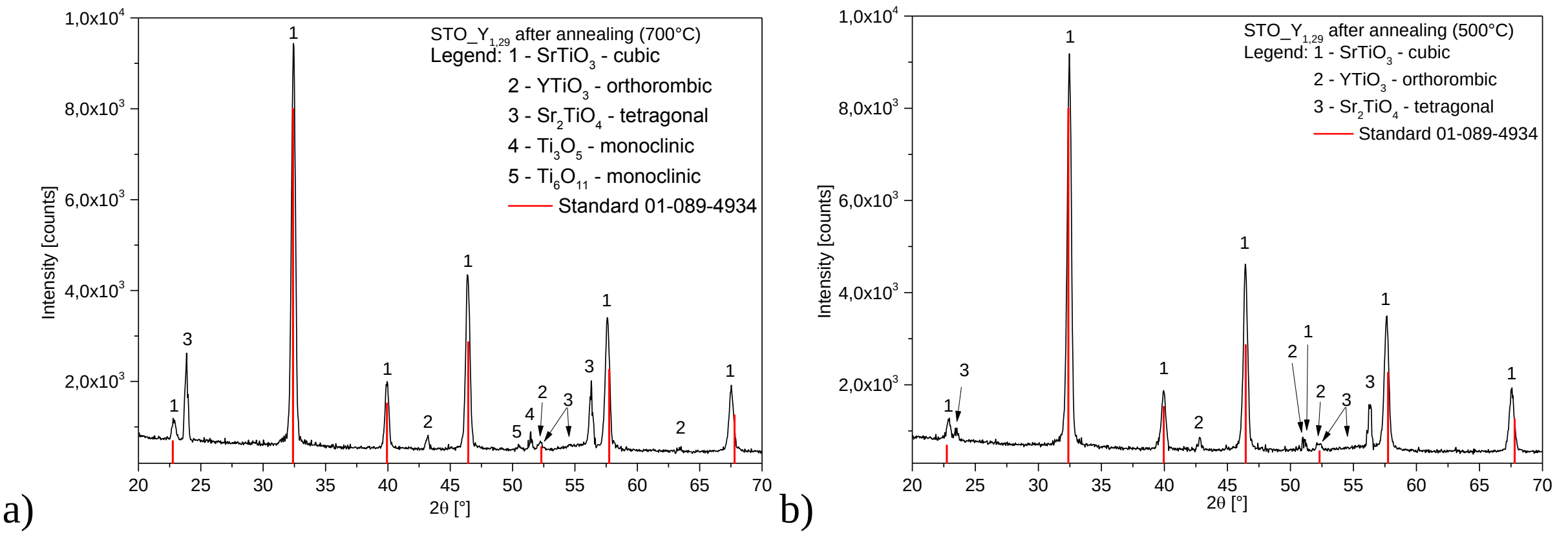

Fig 4 – Diffractograms of the sample STO:$Y_{1,39}$ : (a) annealing temperature 700°C, (b) 500°C. Diffractograms show the difference with lower annealing temperature. The lower temperature was used to reduce the thermal stress in the samples during annealing and subsequent cooling in the HTK.

The third dopant of pure STO used was iron. The conditions of sample production were as in the case of nickel doping. Different amounts of iron pellets were used. A single pellet proved to be the most suitable proportion of the doped element. In this case, the amount of iron in the structure was 1.47 at %.

In the as-deposited state the sample was again amorphous. Two STO phases with different stoichiometry were identified in the diffractogram, both in the initial stage of crystallization. These were the $Sr_3Ti_2O_7$ phase and the $Sr_2Ti_6O_{13}$ phase. Again, the amounts of both phases in the as_deposited state were not determined as they would be burdened with too much error.

A diffractogram of a sample with a single iron pellet after annealing in a high temperature chamber at 500°C is shown and evaluated in Figure 5. The diffraction record shows quite clearly that a full phase transformation from the amorphous phase to the crystalline phase has occurred. Two phases were identified in the sample. The majority phase is the STO phase, which was identified using standard 01-089-4934 from the ICDD - PDF-2 database. This phase has a cubic space lattice with a space group of Pm-3m. This is the identical phase that has already been identified in previous samples. A minority phase was identified as $Ti_3O$, however, this phase is only present in the sample in minimal amounts that were just above the detection limit of the method used.

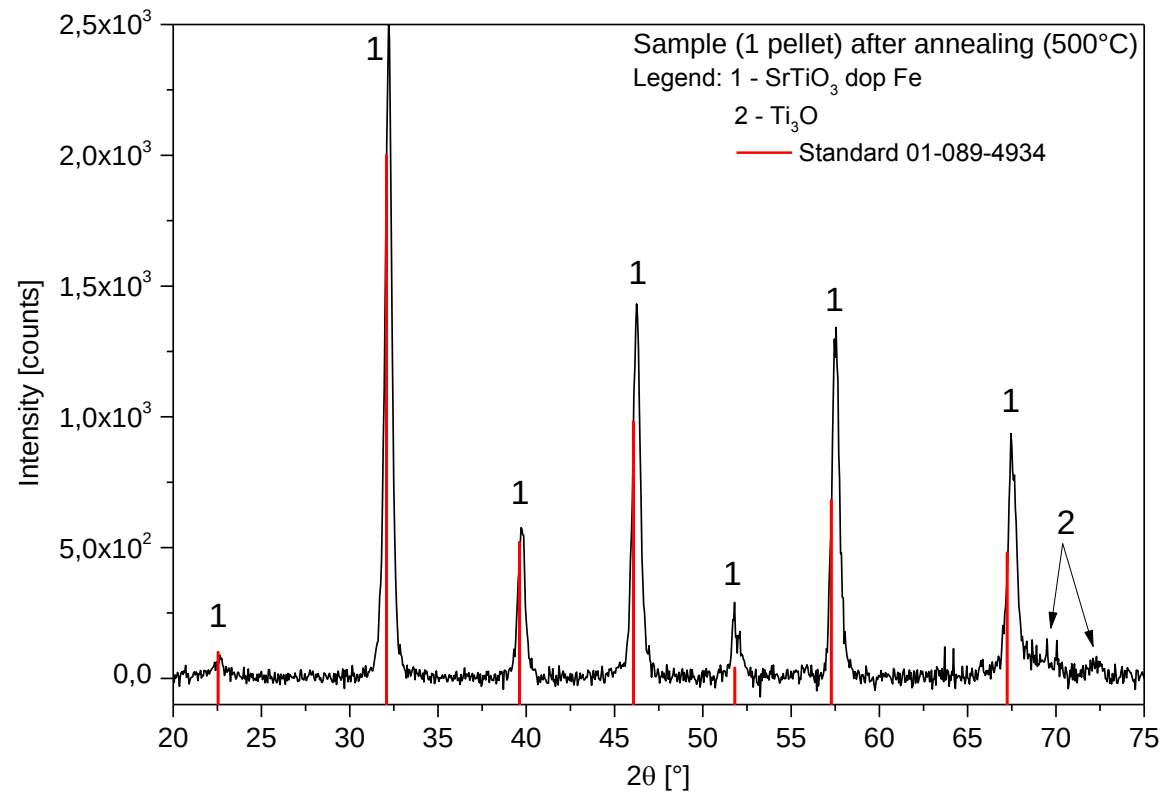

Fig 5 – Diffractogram of STO:$Fe_{1,47}$. There are only two phases, doped STO and $TiO_2$

SEM

Surface examination and topographical data detection by SEM was used on all fabricated thin layers of STO doped with nickel, yttrium and iron. All samples were cleaned in an ultrasonic cleaner before being placed in the SEM vacuum chamber, then immersed in liquid nitrogen for 10 seconds and then fractured. The surface quality of the thin layer, possible surface defects and the fracture edge of the thin layer were investigated for all samples.

The first sample evaluated was sample STO:$Ni_{1.26}$. In Figure 6 a) at 50.000x magnification, it is clearly visible that the surface of the thin layer is unfortunately fractured and thus does not form a continuous surface. From the shape of the cracks, it can be inferred that cracking occurs along the boundaries of some grains. It is further evident from the evaluation of the cracks in Figure 6 b) that these cracks run through the entire thickness of the layer to the phase interface between the substrate and the layer. At the same time, the layer is documented at 100.000x magnification in the figure, showing the layer morphology. It can be seen that it is a fine-grained structure that contains a large number of pores. Both images further show the fact that in this sample there is very good epitaxial growth of the layer on the substrate and thus the layer is well bonded to the substrate. As in all the following samples, the fracture on the substrate is cleavable.

Figure 6 c) and d) show the surface of the sample STO:$Y_{1,39}$. The images show very good epitaxy on the surface of the Si substrate, with no signs of flaking. In the middle of image 6 c) there is a crack running the full thickness of the layer down to the substrate. Furthermore, in image 6 d), at the magnification used of 50.000x, the columnar growth of the layer is visible, corresponding to the sputtering technology used. Next, Figure 6 d) shows the layer at a magnification of 30.000x, which also shows cracking of the sample surface. These cracks run along the boundaries of the individual grains, which are clear from the image. However, the structure of the layer itself and its surface are very good. The surface is smooth with no signs of unevenness or inhomogeneities.

Images of the STO:$Fe_{1,47}$ sample are shown in Figures 6 e) and f). A 45° oblique view at 30.000 x magnification is shown in Figure 6 e). In this image it can be seen that the surface of the thin layer along the grain boundaries has also been cracked. The reasons, implications and possible solutions will be addressed at the conclusion of this paper, including suggesting possible corrections or implementing sputtering procedures for the samples. However, it is evident from the image that there is no flaking of the thin layer and, apart from the cracks mentioned, the layer is compact.

Figure 6 f) highlights a location on the fracture surface that is detailed at 200.000x magnification. Three markers are prominent in this image: it is a fine-grained structure, the fracture of this thin layer is distinctly pitted and tenacious, and the pores in the structure are clearly visible on the fracture surface.

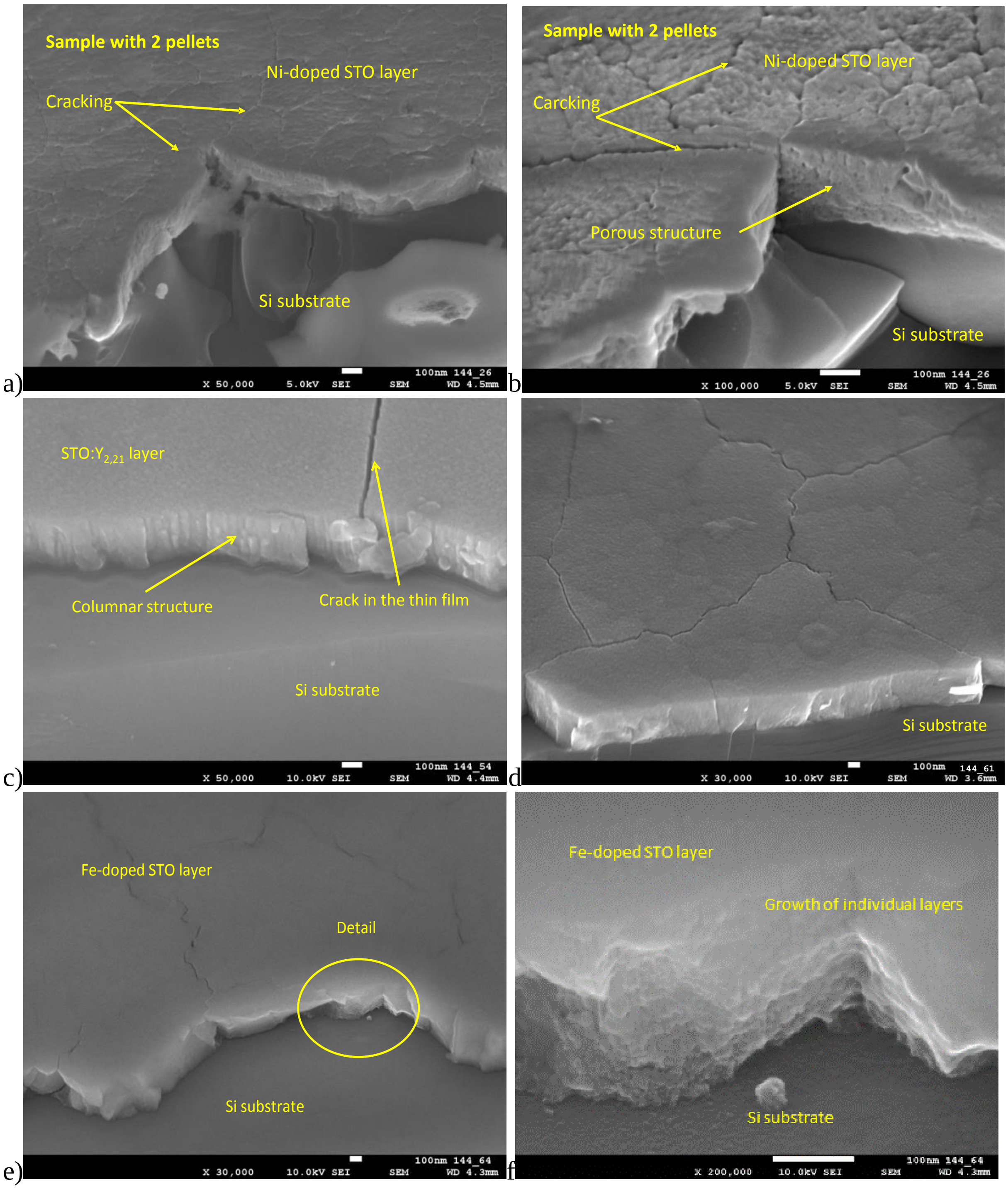

Fig 6 - Images of samples in SEM. Figure 6 (a) shows a sample with two STO:$Ni_{1.26}$ pellets at 50.000x magnification and 6 (b) at 100.000x magnification. The porous layer structure and good adhesion at the interfacial interface can be seen. Figure 6(c) and (d) show the STO:$Y_{1,39}$ sample. Again, good layer structure and columnar direction of layer growth is evident. In Figures 6 e) and f) is a snapshot of the STO:$Fe_{1,47}$ sample. Here the morphology of the layer surface can be seen in image (e) and then the growth direction of the layer and its porous structure can be seen in detail (f). Including the very good interfacial layer-substrate adhesion.

XPS

The evaluated binding energies of the STO:$Ni_{1,26}$ sample are presented in Figure 7(a)-(d), where separate plots of the spectra of the ions (a) O1s, (b) Sr3d, (c) Ti2p and (d) Ni2p are presented in turn. The evaluated plot for oxygen is shown under a). As can be seen, Peak 1 which presents the binding of metal oxides that is tightly bound in the middle of the STO cubic lattice surfaces and whose ions are in the peak positions of the oxygen octahedron has a binding energy of 529.3 eV. From the available data [10], the binding energy for the metal oxide is 529.4 eV. The two values of binding energy are very close to each other and thus it can be assumed that the oxygen ion is not affected by the Ni dopant. The other peaks (Peak 2 and Peak 3) correspond to the binding of oxygen, which is either adsorbed on the sample surface and forms bonds with carbon (Peak 2) or forms single or double bonds with hydrogen (Peak 3). A similar situation is under (b) of the same figure for strontium. Strontium has a valence band in the Sr3d layer where there are two peaks in the doublet. The stronger one with the Sr3d5/2 and Sr3d3/2 labels. The binding energy of Sr3d5/2 corresponds in its value of 132.6 eV to the database value of 132.8 eV, source [11].

The value of the binding energy of the third STO element, titanium Ti2p, is given under (c). As in the case of strontium, this is a doublet, this time Ti2p3/2 (corresponding to the Ti3+ ion) and Ti2p1/2 (Ti4+). The value of the binding energy is 458.1 eV. From [11] it can be found that the binding energy in the STO crystal is 457.4 eV. Thus, it can be seen that the binding energy is affected, namely, by the addition of the dopant Ni, whose spectrum is shown under d) of the identical figure. The Ni2p3/2 position has a binding energy value of 855.4 eV. This value refers to the Ni-O bond. The value found in [12] for Ni2p2/3 alone is 852.7 eV. The match in [13] refers to Ni-O bonding with a binding energy value of 856.0 eV. The bond corresponds to the assumption of Ni doping, where this element substitutes Ti at its position in the middle of the oxygen octahedron, possibly forming direct bonds in the NiO phase.

Research on single crystal samples of doped STO has pinpointed the exact locations where nickel is incorporated into the structure. Figure 8 (top) shows the angle-integrated XPS valence band (VB) spectra for undoped STO, $STO_{0.06}$, and $STO_{0.12}$ films. All spectra are normalized to the highest peak of the valence band at 6 eV. The key features observed in the binding energy range between 3 and 8 eV correspond to O 2p states hybridized with Ti 3d states. For the doped samples, arrows indicate the formation of in-gap states at a binding energy of 2–2.5 eV. A more detailed view of these new states is obtained by subtracting the undoped film's VB spectrum from those of the doped films, as shown in Figure 8 (bottom). The intensity of these states increases with the Ni concentration, confirming that they represent Ni 3d impurity levels within the STO band gap. This serves as definitive evidence that nickel is replacing titanium in the STO lattice.

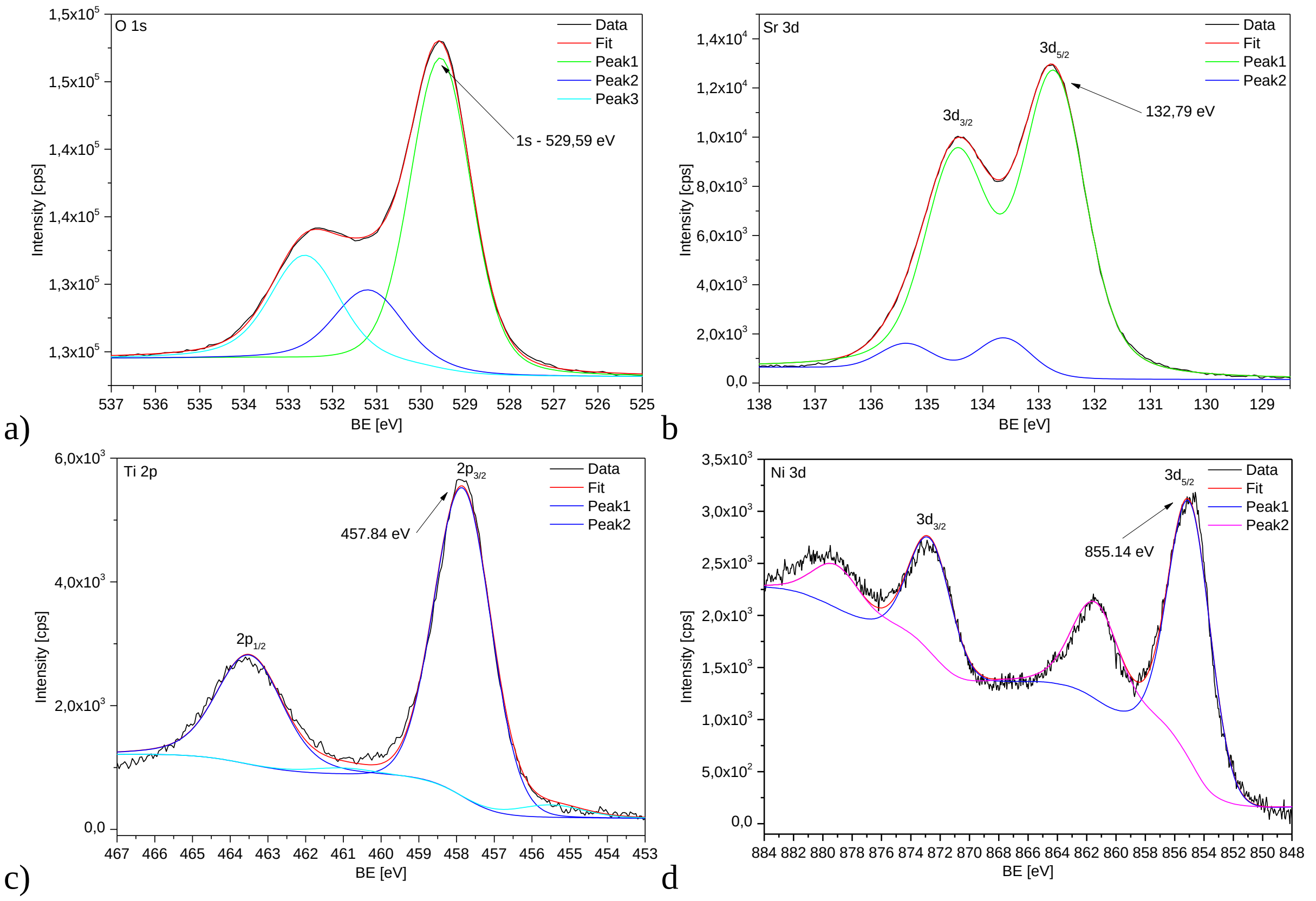

Fig 7 - XPS spectrum of sample STO:Ni$_{1,26}$, beam energy 148.6 eV at room temperature; a) oxygen O1s, b) strontium Sr3d, c) titanium Ti2p and d) nickel Ni2p.

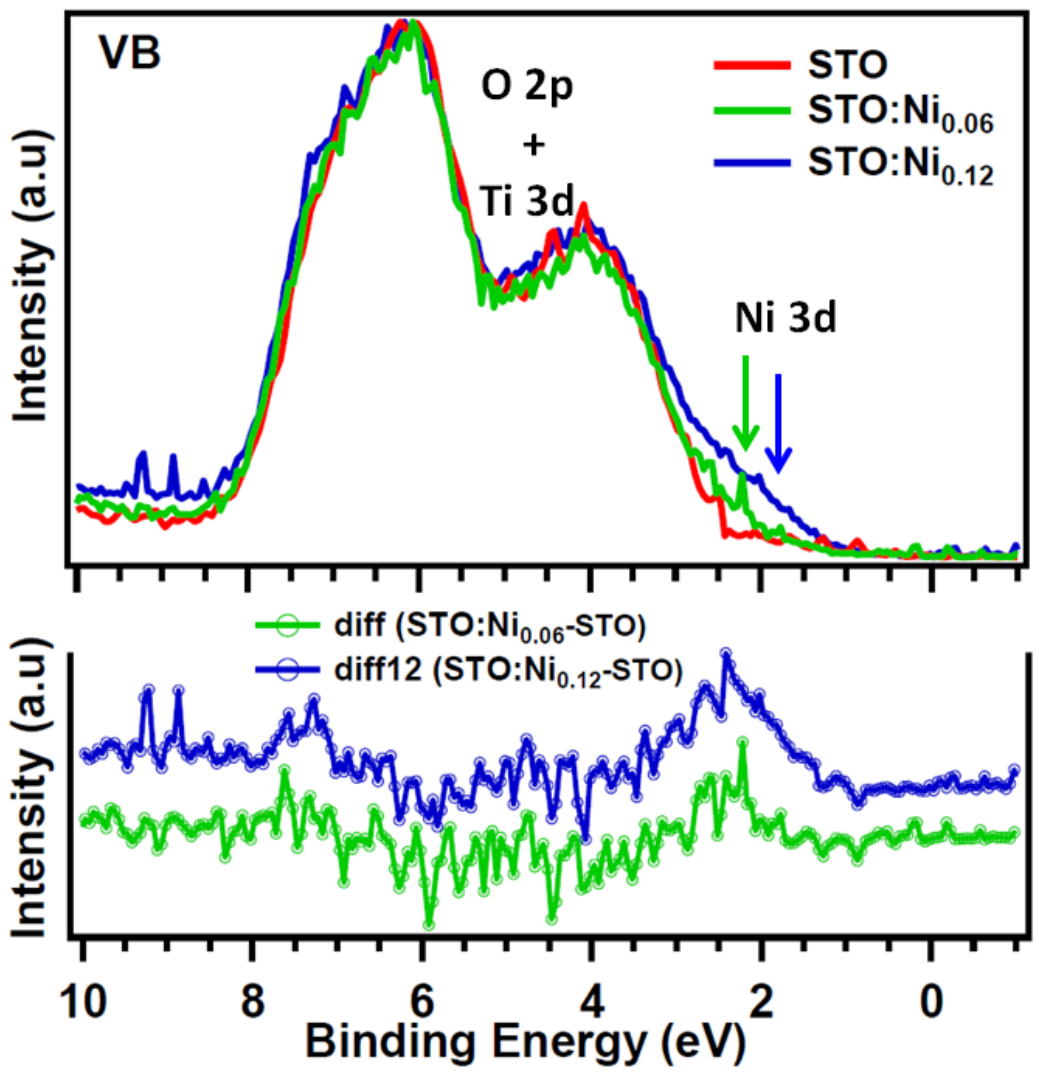

Fig. 8 – Up - The angle-integrated valence band spectra of STO, STO$_{0.06}$, and STO$_{0.12}$ were recorded using monochromatic x-rays with an energy of hν = 1486.6 eV immediately after film growth (in situ). The spectra have been normalized to the valence band's highest peak at 6 eV. Green and blue arrows indicate the presence of in-gap states near the top of the valence band in the doped STO films, appearing around 2–2.5 eV. Down - A quantitative estimation of the new impurity states is given by subtracting the VB of the undoped films from the VB of the doped ones. diff, difference.[14]

Figure 9 a) and b) show the measured values of the sample STO:$Y_{1,39}$ for the basic element titanium a) and its substituent dopant yttrium b), respectively. Figure 9 c) and d) show a similar situation for the sample STO:$Fe_{1,47}$.

In the case of the third basic element titanium Ti2p of sample STO:$Y_{1,39}$, a binding energy value of 457.8 eV was found. Again, this is a doublet of Ti2p3/2 (corresponding to the Ti3+ ion) and Ti2p1/2. The binding energy value is 457.8 eV. From [53] it can be found that the binding energy of Ti2p in the STO crystal is 458.2 eV. Thus, it can be seen that the binding energy is affected and corresponds to the value from [15,16], which is 458.1 eV. From the difference in values, it can be inferred that titanium is affected by the addition of yttrium dopant.

The yttrium dopant element Y3d had a measured binding energy value of 157.1 eV as shown in Figure 9 b). Comparing with the value for metallic Y and the value from [17], which is 157 eV. Due to the shift of the binding energy, it is clear that at this amount (1.39 at %) the dopant diffusely replaces the Ti element in the fundamental lattice more and thus there is a distortion. Also, the differentiation of the two peaks in the Y3d doublet is more pronounced.

The measured values of the basic element and dopant of the STO:$Fe_{1.47}$ sample are shown in Figure 9 c) and d). In the case of Ti2p titanium, the binding energy was found to be 457.9 eV. When compared with the database, it can be seen that the binding energy has been affected as the binding energy value in the STO crystal is 458.2 eV. In [15], it can be traced that the binding energy value of 457.9 eV corresponds to the Ti-O binding energy value, which according to [15] is 458.7 eV. Of course, here too the measured doublet is Ti2p3/2 (corresponding to the Ti3+ ion) and Ti2p1/2.

The iron dopant Fe2p had a measured binding energy of 710.1 eV for sample 144_64 (STO:$Fe_{1,47}$). The binding energy value for metallic Fe is 706.9 eV according to [12]. As can be seen, there was a shift of the binding energy by Fe-O bonding. This binding energy corresponds with its value of 710.3 eV to the Fe-O binding according to [18]. Thus, there was a replacement of the nickel nodal element in the STO base lattice, respectively in the oxygen octahedron of the base lattice. Thus, it is clear that the iron dopant amount of 1.47 at. % causes the diffusive exchange of the Ti basic element.

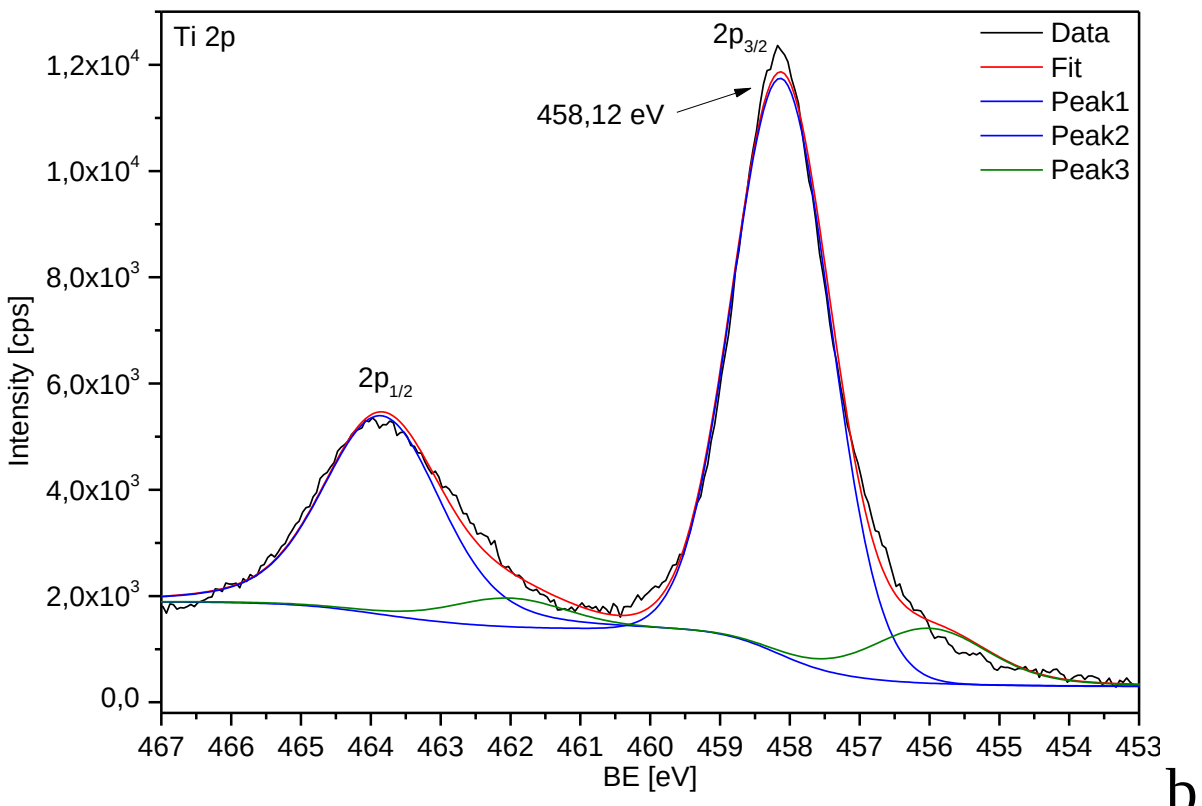

a)

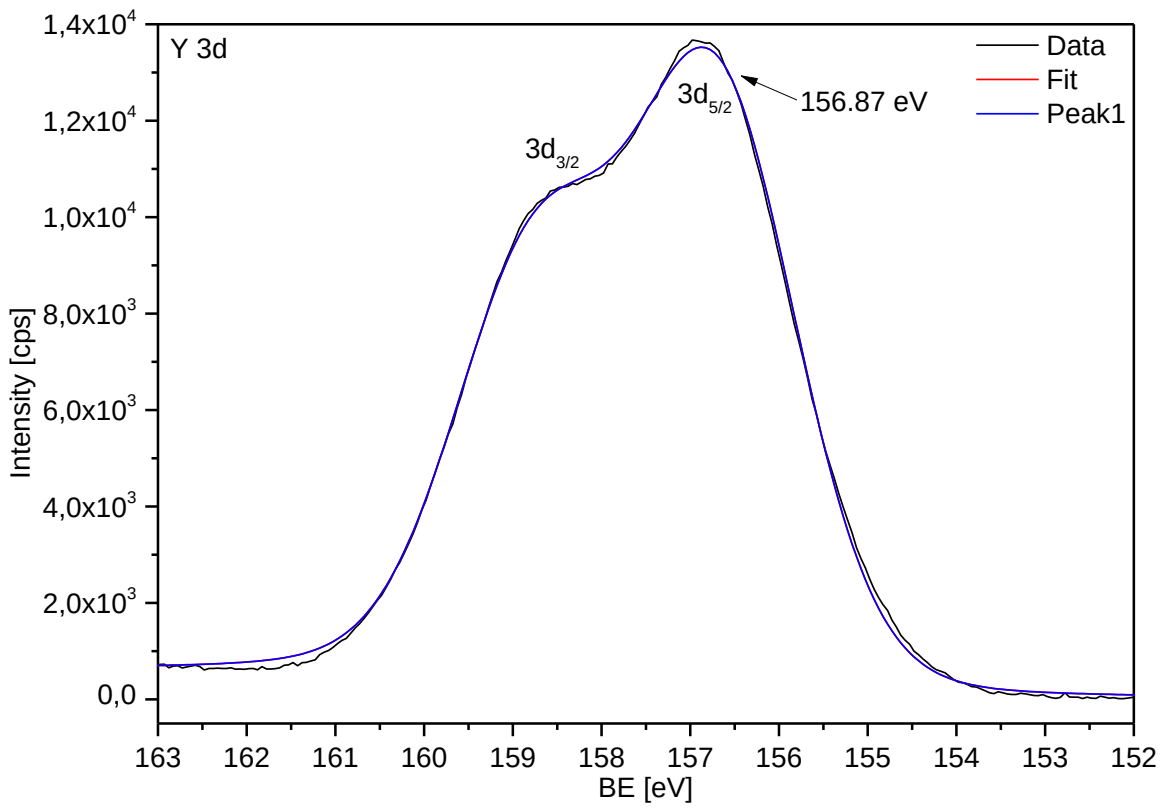

b

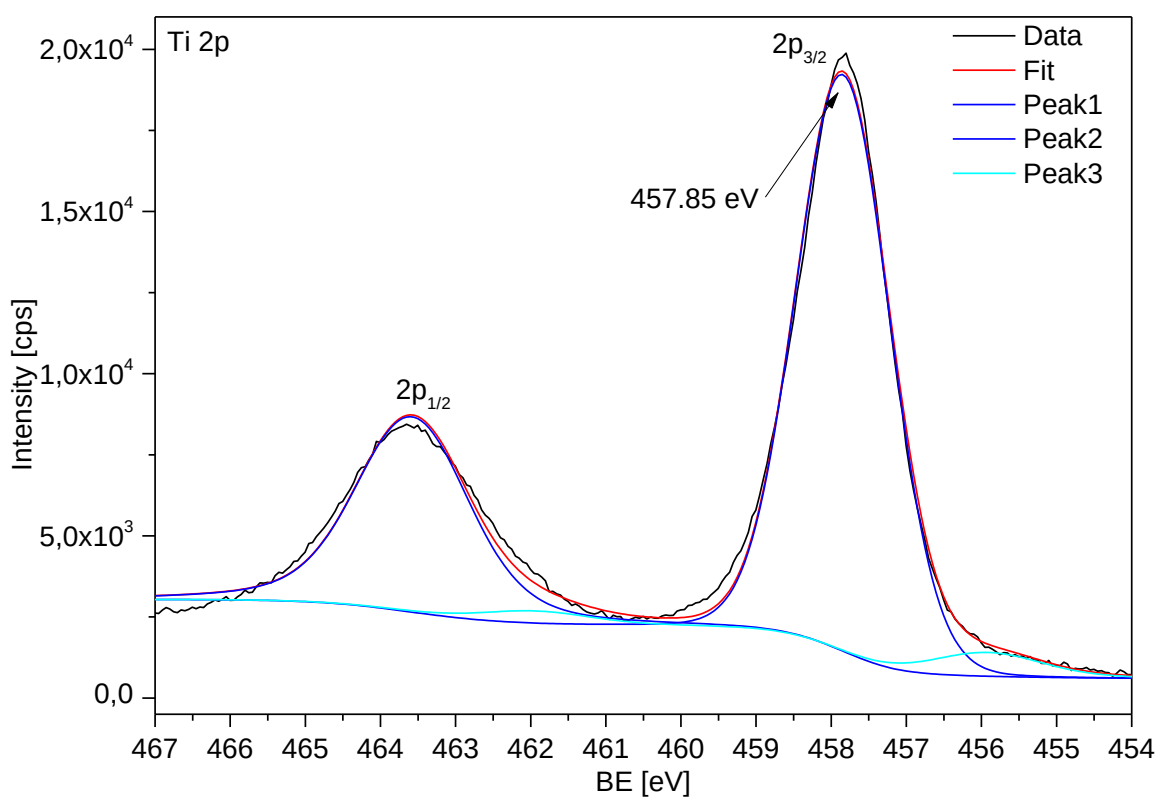

c)

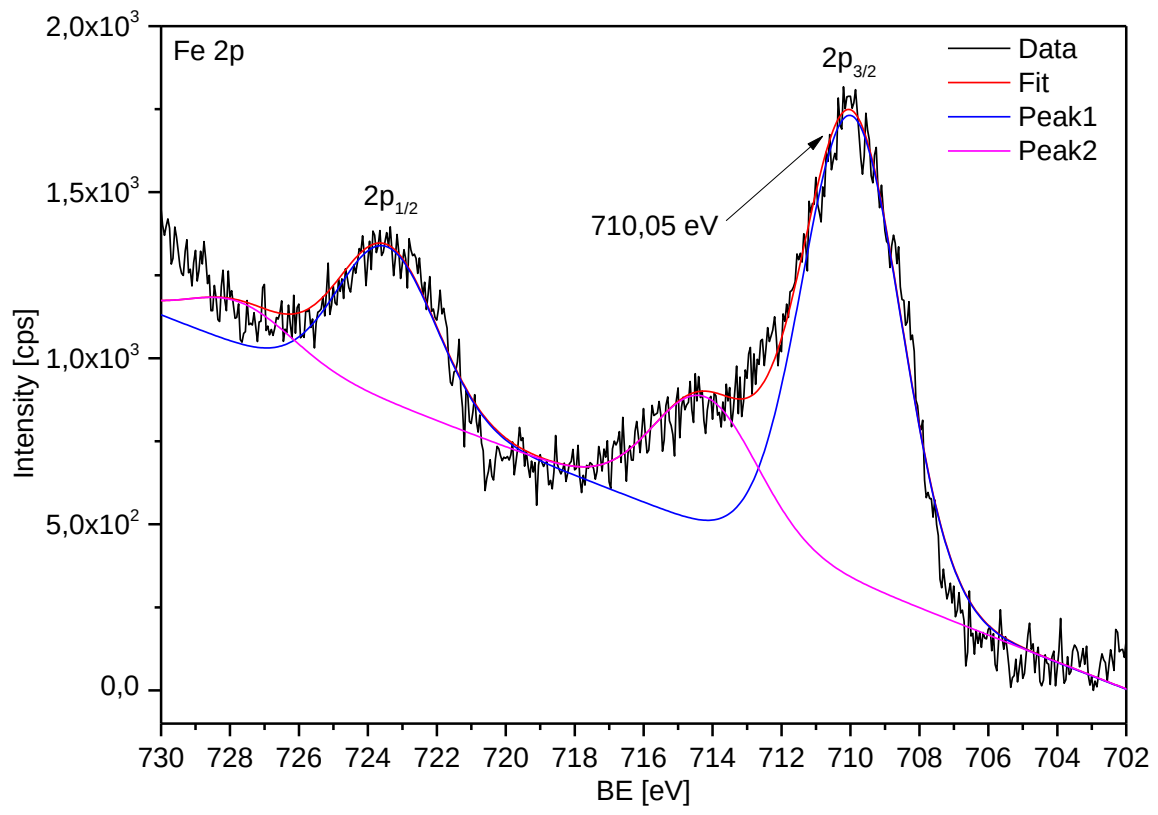

d

Fig 9 - XPS spectrum a) and b) of sample STO:$Y_{1,39}$ a) – Ti2p, b) – Y3d, and picter c) and d) of sample STO:$Fe_{1,47}$ c) – Ti2p, d) – Fe2p; beam energy 148.6 eV at room temperature.

## 4. Conclusions

The conclusion of this thesis evaluates whether the primary objectives have been met. Based on the analyzed data, all samples exhibited notable changes in both structural and electronic properties due to the doping of Ni, Y, and Fe elements, as compared to the undoped STO state. Each dopant positively influenced the structural properties, refining the thin film's structure and leading to the formation of voids and pores throughout. This successfully verifies the objective of assessing the impact of the dopants. These pores play a crucial role in the suitability of the thin films for photocatalytic applications, where chemical reactions, such as the decomposition of $H_2O$ into $H_2$ and O, take place [19,20,21,22]. Other potential applications include biofuel production, as previously mentioned [23].

This work also determined the phase transformation temperature from the amorphous phase, created after magnetron sputtering at room temperature, to the crystalline phase. This secondary objective arose from addressing the cracks observed in the thin films' surface, which made conducting electrical tests, such as those using the Hall effect, infeasible. Initially, an annealing temperature of 800°C was used, but further experimentation revealed the phase transformation temperature to be between 460-480°C. Despite this, the cracking issue persisted, a problem also reported in [20]. In that study, STO thin films on silicon substrates, prepared via the sol-gel method, exhibited similar cracking. The probable causes were attributed to the critical thin film thickness of 150 nm and the visco-elastic behavior of the film during heat treatment and relaxation. These findings align with the experimental results of this work, where surface cracking occurred post-annealing due to relaxation. Consequently, a minimum film thickness of 150 nm, already in the crystalline phase upon deposition, is recommended to avoid post-deposition thermal processing.

Furthermore, the experiment included measuring the binding energies of the STO base lattice and the dopants. These values were discussed in detail in the relevant chapters. In the discussion section, I evaluate the expected influence of the titanium ion Ti2p position. In this work, we have demonstrated a shift in the binding energy of the titanium ion in all samples evaluated compared to the undoped STO base lattice. These shifts are attributed to the incorporation of the doping elements (Ni, Y, Fe) into the lattice. The other two dopants showed minimal impact on binding energy, on the order of hundredths of an eV. These results align with similar research, such as [24], which also observed a reduction in the Eg forbidden bandwidth from 3.2 eV to 1.8 eV.

Acknowledgement

This publication was supported by the project Quantum materials for applications in sustainable technologies (QM4ST), funded as project No. CZ.02.01.01/00/22_008/0004572 by Programme Johannes Amos Commenius, call Excellent Research.